\documentclass[twocolumn,showpacs,amsmath,amssymb,floatfix]{revtex4}
\usepackage{graphicx}% Include figure files
\usepackage{dcolumn}% Align table columns on decimal point
\usepackage{bm}% bold math
\usepackage{textcomp}
\usepackage{epsfig}
\usepackage{color}
\newcommand{\beq}{\begin{equation}}
\newcommand{\eeq}{\end{equation}}
\newcommand{\bea}{\begin{eqnarray}}
\newcommand{\eea}{\end{eqnarray}}

\newcommand{\bary}{\begin{array}}
\newcommand{\eary}{\end{array}}
\newcommand{\benum}{\begin{enumerate}}
\newcommand{\eenum}{\end{enumerate}}
\newcommand{\bitem}{\begin{itemize}}
\newcommand{\eitem}{\end{itemize}}
\begin{document}
\title{Gauge invariance in the theoretical description of time-resolved angle-resolved pump/probe photoemission spectroscopy}

\author{J. K. Freericks$^1$,
H. R. Krishnamurthy$^{2,3}$,
M.A. Sentef$^4$, and T.P. Devereaux$^{4,5}$}
\affiliation{$^1$Department of Physics, Georgetown University, 37th and O Sts. NW, Washington, DC 20057, USA\\
$^2$Centre for Condensed Matter Theory, Department of Physics, Indian Institute of Science, Bangalore 560012, India\\
$^3$Jawaharlal Nehru Centre for Advanced Scientific Research, Bangalore 560064, India\\
$^4$Stanford Institute for Materials and Energy Science, SLAC National Accelerator Laboratory, Menlo Park, CA 94025, USA\\
$^5$Geballe Laboratory for Advanced Materials, Stanford University, Stanford, CA 94305, USA
}
\date{\today}

\begin{abstract}
Nonequilibrium calculations in the presence of an electric field are usually performed in a gauge, and need to be transformed to reveal the gauge-invariant observables. In this work, we discuss the issue of gauge invariance in the context of time-resolved angle-resolved
pump/probe photoemission. If the probe is applied while the pump is still on, one must ensure that the calculations of the observed photocurrent are gauge invariant. We also discuss the requirement of the photoemission signal to be positive and the relationship of this constraint to gauge invariance.  We end by discussing some technical details related to the perturbative derivation of the photoemission spectra, which involve processes where the pump pulse photoexcites electrons due to nonequilibrium effects.
\end{abstract}
\maketitle

\section{Introduction}

Recently,  the framework for a general theory of time-resolved photoemission was developed~\cite{our-prl}, where a system is pumped using a high power laser (but typically one whose photons do not have enough energy to photoexcite electrons) into an excited (non-equilibrium) state, and then probed using another (relatively low intensity) laser pulse whose photons do have enough energy to photoexcite, after a controlled and variable time delay. In most cases, the pump laser is turned off before the probe laser is turned on, and the previous theory detailed precisely what (non-equilibrium) correlation function of the system is measured in such an experiment~\cite{our-prl}. The result was determined to leading (second) order in the probe Hamiltonian $(\mathcal{H}_{\rm probe})$, and some approximations that can be used to simplify its calculation were also discussed.

Here, we extend that previous work to the cases where the pump pulse continues to be on when the probe pulse becomes operative. In such cases, although the framework developed in Ref.~\onlinecite{our-prl} continues to be valid, several expressions given there cannot be used because they are not general enough to ensure manifest gauge invariance and to take into account all of the required time dependence; hence they can lead to erroneous results, and it becomes necessary to employ instead the expressions presented here. This issue motivates us to discuss more thoroughly the nature of gauge invariance in pump/probe photoemission, where we relate it to the condition that the measured response function must be nonnegative. To be concrete, we examine this situation for noninteracting band electrons, where explicit formulas can be developed, and the relationship between gauge invariance and constraints on the measured signal become clear.
 
Furthermore, the discussions in Ref.~\onlinecite{our-prl},  neglected some additional contributions to the detected photocurrent $\langle \mathbf{J}_d \rangle(t)$ that are formally also of second order in the probe Hamiltonian $\mathcal{O}[(\mathcal{H}_{\rm probe})^2]$, simply stating that the term kept makes the most {\em dominant} contribution. Here we show explicitly why those other contributions are indeed small, when compared to the term that is traditionally kept.

%\begin{itemize}

%\item

\section{Gauge Invariance issues}

The procedure to determine the photocurrent is completely straightforward. We start by introducing the field via a Peierls substitution and evolve the system with an evolution operator $U(t_1,t_2)$ that includes the time-dependent effects of the field [with the Hamiltonian that includes the effects of the time-dependent pump field denoted by $\mathcal{H}_{\rm pump}(t)$ in the Schroedinger representation]. Then we turn on a weak probe Hamiltonian $\mathcal{H}_{\rm probe}(t)$ which is responsible for the photoemission.  The photocurrent operator representing the
detector, which is  designed to detect photoelectrons with momentum {\bf k} peaked around ${\bf k}_e$
and localized at the detector position ${\bf R}_d$ outside the sample, is 
\begin{equation}
{\bf J}_d=\frac{\hbar {\bf k}_e}{m_e}c^\dagger_{{\bf k}_e;{\bf R}_d}c^{}_{{\bf k}_e;{\bf R}_d}
\end{equation}
where $c^\dagger_{{\bf k}_e;{\bf R}_d}$ creates an electron in a wave-packet state with a
momentum space wave function that is both strongly peaked around the momentum value ${\bf k}_e$ and also peaked around the spatial location ${\bf R}_d$ of the detector.

As mentioned above, even in contexts where the pump pulse continues to be present when the probe pulse is on, the initial part of the discussion in Ref.~\onlinecite{our-prl} leading up to the expression in its Eq. (2) for the measured photocurrent, namely,
\bea
\langle \mathbf{J}_d \rangle(t) = \frac{1}{(\hbar)^2} \int_{t_0} ^{t} dt_2 \int_{t_0} ^{t} dt_1 \langle U(-\infty, t_2)\mathcal{H}_{\rm probe}(t_2)  \nonumber\\
\times \; U(t_2,t) \mathbf{J}_d  U(t,t_1) \mathcal{H}_{\rm probe}(t_1) U(t_1, -\infty) \rangle_{\mathcal{H}};
\label{J_d-avg-detail}\\
\langle O \rangle_\mathcal{H}  \equiv \sum_n \rho_n \langle {\Psi}_n \left |O \right | \Psi_n\rangle = \mathcal{Z}^{-1}{\rm Tr}[e^{-\mathcal{H}/(k_B T)} O].\,\nonumber
\eea 
continues to be valid; here, $\mathcal{H}$ is the equilibrium Hamiltonian with no field, $\mathcal{Z}$ is its corresponding partition function, and the evolution operator $U$ evolves with respect to the Hamiltonian with the pump field $\mathcal{H}_{\rm pump}(t)$. The probe Hamiltonian [in the Heisenberg representationwith respect to $\mathcal{H}_{\rm pump}(t)$] $U^\dagger(t_1,-\infty)\mathcal{H}_{\rm probe}(t_1)U(t_1,-\infty)$, now has an additional time dependence due to the time dependence of the vector potential of the pump pulse. Hence Eq. (3) of Ref.~\onlinecite{our-prl} for the component of the probe Hamiltonian responsible for the the absorption of a photon of momentum $\hbar\mathbf{q}$ and the ejection of an electron from $\nu \mathbf{k}_{\parallel}$ to $\nu^\prime \mathbf{k}_{\parallel} + \mathbf{q}_{\parallel}$, (where $\mathbf{k}_{\parallel}, \mathbf{k}_{\parallel} + \mathbf{q}_{\parallel}$ label the electron wave vector components parallel to the surface, and $\nu, \nu^\prime$ the other indices  or quantum numbers specifying the one electron band states of the sample in the presence of a plane surface,) needs to be appropriately modified, and rewritten as
\bea
&~&\mathcal{H}_{\rm probe}(t_1)=\label{PES-ham}\\
&~&\sum_{\nu,\nu^\prime, \mathbf{k}_\parallel} s(t_1) e^{i \omega_{\mathbf{q}} t_1 }  M_{\mathbf{q}}(\nu,\nu^\prime; \mathbf{k}_\parallel; t_1) c^{\dagger}_{\nu^\prime \mathbf{k}_{\parallel} + \mathbf{q}_\parallel} c_{\nu \mathbf{k}_\parallel} a_{\mathbf{q}} .\nonumber
\eea
with the matrix-element associated with the above process being replaced by its time dependent version, $M_{\bf q}(\nu,\nu^\prime;{\bf k}_\parallel)\rightarrow M_{\mathbf{q}}(\nu,\nu^\prime; \mathbf{k}_\parallel; t_1 )$, given by the {\em modified} expression
\bea
&~&M_{\mathbf{q}}(\nu,\nu^\prime; \mathbf{k}_\parallel; t_1 ) =\label{mat-element}\\
&~& \langle \nu^\prime \mathbf{k}^\prime_\parallel  |\frac{ ie \hbar {\mathbf{A}}_{probe}(\mathbf{r},t_1)}{m_e c}\cdot \left [\nabla -\frac{ i e  {\mathbf{A}}_{pump}(\mathbf{r}, t_1)}{\hbar c}\right ] | \nu \mathbf{k}_\parallel \rangle, \nonumber
\eea
which depends on the vector potential of the pump field and hence inherits its additional time dependence. The symbol $\omega_{\bf q}=c|{\bf q}|$ is the photon frequency of the photons in the probe pulse.

Plugging the expression for this matrix element [Eq.~(\ref{mat-element})] into the formula for the probe Hamiltonian in Eq.~(\ref{PES-ham}), and then into the photocurrent expectation value in Eq.~(\ref{J_d-avg-detail}), and extracting the contribution of the photocurrent that has momentum ${\bf k}_e$ at position ${\bf R}_d$, then yields the total number of photoelectrons emitted from the sample and detected at the detector for all times between time $t$ and time $t_0$, which we call $P_{\bf k}(t)$.  The correct expression becomes
\bea
P_{\bf k}(t) \equiv  \frac{1}{\hbar^2} \sum_{\nu_1,\nu_1^\prime, \mathbf{k}_{\parallel 1}}  \sum_{\nu_2,\nu_2^\prime, \mathbf{k}_{\parallel 2}} \int_{t_0}^{t} dt_2 \int_{t_0}^{t} dt_1 s(t_2)s(t_1)  \nonumber  \\
\times  e^{i\omega_{\mathbf{q}}(t_1-t_2)} M^*_{\mathbf{q}}(\nu_2,\nu_2^\prime; \mathbf{k}_{\parallel 2}; t_2) \, M_{\mathbf{q}}(\nu_1,\nu_1^\prime; \mathbf{k}_{\parallel 1}; t_1) \,  \; \nonumber \\
 \times \;  \langle c^{\dagger}_{\nu_2 \mathbf{k}_{\parallel 2}}(t_2) c_{\nu_2^\prime \mathbf{k}_{\parallel 2}+\mathbf{q}_\parallel}(t_2) c^{\dagger}_{\nu^\prime \mathbf{k}_\parallel^\prime}(t) c_{\nu \mathbf{k}_\parallel}(t)  \nonumber\\
\times c^{\dagger}_{\nu_1^\prime \mathbf{k}_{\parallel 1}+\mathbf{q}_\parallel}(t_1) c_{\nu_1 \mathbf{k}_{\parallel 1}}(t_1) \rangle_{\mathcal{H}}. \; \;
\label{central-result-2}
\eea
In most experimental contexts the photoejected electrons have a high enough energy that their propagation is uncorrelated with the other (lower energy) electronic excitations of the system. Under these conditions, the three particle current correlation function in Eq.~(\ref{central-result-2}) given by the six operator average can be factorized as:
\bea
\langle c^{\dagger}_{\nu_2 \mathbf{k}_{\parallel 2}}(t_2) c_{\nu_1 \mathbf{k}_{\parallel 1}}(t_1)\rangle_{\mathcal{H}} \; \times \; \langle c_{\nu_2^\prime \mathbf{k}_{\parallel 2}+\mathbf{q}_\parallel}(t_2) c^{\dagger}_{\nu^\prime \mathbf{k}_\parallel^\prime}(t)\rangle_{\mathcal{H}}  \; \nonumber\\
\times \langle c_{\nu \mathbf{k}_\parallel}(t)  c^{\dagger}_{\nu_1^\prime \mathbf{k}_{\parallel 1}+\mathbf{q}_\parallel}(t_1)\rangle_{\mathcal{H}} \simeq \nonumber\\
\langle c^{\dagger}_{\nu_2 \mathbf{k}_{\parallel 2}}(t_2) c_{\nu_1 \mathbf{k}_{\parallel 1}}(t_1)\rangle_\mathcal{H} \, \delta_{\nu_2^\prime, \nu^\prime}  \delta_{\nu_1^\prime, \nu} \delta_{\mathbf{k}_{\parallel 2}+\mathbf{q}_\parallel, \mathbf{k}_\parallel^\prime} \nonumber \\
  \delta_{\mathbf{k}_{\parallel 1}+\mathbf{q}_\parallel, \mathbf{k}_\parallel}
e^{i [(\epsilon_{\nu^\prime \mathbf{k}_\parallel^\prime}-\mu)(t-t_2)- (\epsilon_{\nu \mathbf{k}_\parallel} - \mu)(t-t_1)]/\hbar}.
\label{fac-eqn}
\eea
Now, if the pump pulse is on when the probe pulse ejects the photoelectron, there is a further approximation involved in going from the LHS to the RHS of Eq.~(\ref{fac-eqn}), in that the effect of the pump-pulse on the propagation of the photoexcited electron (contained in the two averages with respect to $\mathcal{H}$) has also been neglected; this then ignores effects like the ponderamotive force acting on the photoexcited electrons. Given this approximation, Eq. (7) of Ref. \onlinecite{our-prl} for $P_{\bf k}(t)$ has now to be modified as follows:
\bea
P_{\bf k}(t) \simeq -i \frac{1}{\hbar^2} \sum_{\nu_1,\nu_2} \int_{t_0} ^{t} dt_2 \int_{t_0} ^{t} dt_1 s(t_2) s(t_1) e^{i\omega(t_1- t_2)}  \times \nonumber \\
 M^*_{\mathbf{q}}(\nu_2,\nu_e; \mathbf{k}_{e\parallel}; t_2) M_{\mathbf{q}}(\nu_1,\nu_e; \mathbf{k}_{e\parallel }; t_1) G^<_{\nu_1 \mathbf{k}_{e\parallel}, \nu_2 \mathbf{k}_{e\parallel}}(t_1, t_2),
\label{P-to-G-lesser}
\eea

We note that in contexts where the pump pulse is turned off before the probe pulse is turned on, as was the case in all the detailed calculations reported and discussed in Ref.~\onlinecite{our-prl}, the matrix elements $M$ no longer have any time dependence, and all of the new expressions described above reduce to the ones given previously. However, if the pump pulse is present when the probe pulse is on, then the above expressions are relevant, and it is to be expected that the effect of the vector potential of the pump field on the matrix elements $M$ as well as on the propagation of the photoexcited electron in Eq.~(\ref{fac-eqn}) has to be correctly taken into account to ensure the independence of $P_{\bf k}(t)$ on gauge transformations of the vector potential of the pump field.

It is customary in calculations of PES of layered systems, especially in those that use model Hamiltonians restricted to a single band and focus on many body effects rather than on band structure effects, not to calculate the matrix elements, but instead to replace them by constants, make the approximation in Eq.~(\ref{fac-eqn}), and furthermore ignore the indices $\nu_1$ and $\nu_2$ to focus on the effects of a single band only. If this approximation is used for calculating $P_{\bf k}(t)$ as given by the modified Eq.~(\ref{P-to-G-lesser}), then the nonequilibrium lesser Green's function in Eq.~(\ref{P-to-G-lesser}), restricted to a single band, and given by 
\beq
G^<_{\mathbf{k}_{e\parallel}}(t_1, t_2)
\equiv i \langle c_{ \mathbf{k}_{e\parallel}}^{\dag} (t_2)  c^{}_{\mathbf{k}_{e\parallel}}(t_1)\rangle_{\mathcal{H}}
\eeq
and calculated in the presence of the pump field, and hence $P_{\bf k}(t)$ itself, will be gauge dependent and incorrect.  However, if one calculates the total photoemission response, using the approximation of a constant matrix element, then the photoemission response is local and gauge-invariant, and one can use the formulas already discussed in Ref.~\onlinecite{our-prl}. It is the angle-resolved photoemission that needs to be corrected.

One way to fix this problem is to follow the prescription by Bertoncini and Jauho \cite{jauho1991} who discovered a constructive transformation that creates a gauge-invariant Green's function. This procedure replaces the lesser Green's function by its gauge invariant modification in the formula for $P_{\bf k}(t)$:
\beq
G^<_{\mathbf{k}_{e\parallel}}(t_1, t_2) \rightarrow {\tilde{G}}^<_{\mathbf{k}_{e\parallel}}(t_1, t_2)  \equiv  G^<_{{\bar{\mathbf{k}}}_{e\parallel}}(t_1, t_2),
\eeq
with
\beq
{\bar{\mathbf{k}}}_{e\parallel} = {\mathbf{k}}_{e\parallel} +\frac{e}{\hbar c} \frac{1}{t_1 - t_2} \int_{-(t_1 - t_2)/2}^{(t_1 - t_2)/2} dt^\prime
{\mathbf{A}}_{pump}\left ( \frac{t_1+t_2}{2}+t^\prime \right )  \,
\eeq
where for simplicity, we have assumed that the spatial dependence of the vector potential of the pump field can be neglected (an approximation which is reasonable for optical or infrared pump fields), and to the extent that the Green's function depends only on momenta parallel to the sample surface, only the parallel component of ${\mathbf{A}}_{pump}$ matters. Another way of stating this is that the momentum
is shifted by the average vector potential for the time interval in the Green's function. This approach has been used in calculations of time-resolved ARPES for electron-phonon interacting systems~\cite{lex1,prx} and in the generation of transient Haldane phases in graphene~\cite{graphene}.

The task of either generalizing this fix, or {\em deriving} the appropriate gauge independent prescription from first-principles, in the contexts where one takes into account the effects of the surface, the effects of three dimensional band structures with mutliple bands crossing the Fermi level, and the effect of the pump pulse on the propagation of the photoelectrons shortly after being photoemitted, {\it etc.}, poses a major theoretical challenge that we do not solve here, and leave for future work.

\section{Gauge invariance and positivity of the angle-resolved photocurrent}

The general formula for the time-resolved and angle-resolved 
photoemission spectra involves the square of matrix elements, and hence should be manifestly
nonnegative.  This is physically important because the photoemission spectrum cannot be negative, as it is a probability.
If we use the standard approximation of replacing the matrix elements in Eq.~(\ref{P-to-G-lesser}) by constants and focusing on a single band for the photoemission, then the expression for the angle-resolved photoemission probability becomes
\begin{eqnarray}
\tilde P_{\bf k}(t)\propto -\frac{i}{\hbar^2}\int_{t_0}^t dt_1\int_{t_0}^t dt_2
s(t_1)s(t_2)
e^{i\omega(t_1-t_2)}\tilde G^<_{\bf k}(t_1,t_2).\nonumber\\
\label{eq: prob_gauge}
\end{eqnarray}
Note that because $\tilde G^<_{\bf k}(t_1,t_2)=-\tilde G^{<*}_{\bf k}(t_2,t_1)$, one immediately establishes that the probability is real by simply interchanging the dummy integration variables $t_1\leftrightarrow t_2$, which shows $\tilde P_{\bf k}=\tilde P_{\bf k}^*$.
If we examine the photocurrent probability in a gauge, where we replace $\tilde G^<$ by $G^<$, then it is easy to prove that the signal is nonnegative. Recalling that 
\begin{equation}
-iG^<_{\bf k}(t_1,t_2)=\langle c^\dagger_{\bf k}(t_2)c^{}_{\bf k}(t_1)\rangle_{\mathcal{H}}
\end{equation}
where the angle brackets denote the trace over states weighted by the initial equilibrium density matrix  and the operators are in the Heisenberg representation with respect to the Hamiltonian with the pump $\mathcal{H}_{\rm pump}(t)$. Then one simply writes
\begin{equation}
P_{\bf k}(t)=\frac{1}{\hbar^2}\sum_n \frac{e^{-\beta E_n}}{\mathcal{Z}}\left | \int_{t_0}^t dt_1 s(t_1)e^{i\omega t_1} c_{\bf k}(t_1)|n\rangle \right |^2,
\end{equation}
which is manifestly nonnegative because the norm of a vector is nonnegative as is the exponential. Note that we use the notation for energy eigenstates $\mathcal{H}|n\rangle=E_n|n\rangle$ for the initial system when it is in equilibrium and we denote $\beta=1/(k_BT)$.

However, an important question to resolve is whether the use of the gauge-invariant Green's function in Eq.~(\ref{eq: prob_gauge}) leads to a nonnegative tr-ARPES signal. Written out in detail, the gauge-invariant time-resolved angle-resolved photoemission spectra is determined by
\begin{widetext}
\begin{equation}
\tilde P_{\bf k}(t)=\frac{1}{\hbar^2}\int_{t_0}^t dt_1 \int_{t_0}^t dt_2
s(t_1)s(t_2)e^{i\omega(t_1-t_2)}
\Big\langle c^\dagger_{{\bf k}+\frac{e}{\hbar c}\frac{1}{t_1-t_2}\int_{t_2}^{t_1}dt'{\bf A}_{\rm pump}\left (t'\right )}(t_2)
c^{}_{{\bf k}+\frac{e}{\hbar c}\frac{1}{t_1-t_2}\int_{t_2}^{t_1}dt'{\bf A}_{\rm pump}\left (t'\right )}(t_1)\Big\rangle.
\end{equation}
\end{widetext}
Because the shift of the momentum is not a function of $t_1$ only for the $c_{\bf k}$ operator and of $t_2$ only for the $c^\dagger_{\bf k}$ operator, the argument used above to show nonegativity of the function in a gauge no longer goes through. In the general interacting case, it is difficult to manipulate these expressions further
because they can have complicated time dependence. Instead, we focus on a concrete example which
can be solved exactly: the noninteracting problem.

\section{Positivity of the angle-resolved photocurrent for a noninteracting single band}

The lesser Green's function in the vector-potential-only gauge for a noninteracting particle satisfies~\cite{bloch}
\begin{equation}
G^<_{\bf k}(t_1,t_2)=if(\epsilon_{\bf k}-\mu)e^{-\frac{i}{\hbar}\int_{t_2}^{t_1} dt' [\epsilon_{{\bf k}-e{\bf A}(t')/(\hbar c)}-\mu]}
\end{equation}
with $f(x)=1/[1+\exp(\beta x)]$ the Fermi-Dirac distribution function. Using this result for the lesser Green's function in the photoemission probability calculated in the gauge yields
\begin{widetext}
\beq
P_{\bf k}(t)=\frac{1}{\hbar^2}f(\epsilon_{\bf k}-\mu)\int_{t_0}^{t}dt_1 
\int_{t_0}^{t}dt_2 s(t_1)s(t_2)
\exp\left [ \frac{i}{\hbar}\int_{t_2}^{t_1}dt'\left ( \hbar\omega+\mu-\epsilon_{{\bf k}-e{\bf A}(t')/(\hbar c)}\right )\right ].
\eeq
As before, one can immediately show that this expression is nonnegative, by writing it as
\beq
P_{\bf k}(t)=\frac{1}{\hbar^2}f(\epsilon_{\bf k}-\mu)\Biggr |\int_{t_0}^{t}dt_1 s(t_1)
\exp \left [\frac{i}{\hbar}\int_{t_0}^{t_1}dt'\left (\hbar\omega+\mu-\epsilon_{{\bf k}-e{\bf A}(t')/(\hbar c)}\right )\right ] \Biggr |^2.
\eeq
The gauge-invariant prescription for the photoemission, however, leads to a complicated expression given by
\bea
&~&\tilde P_{\bf k}(t)=\frac{1}{\hbar^2}\int_{t_0}^{t}dt_1 
\int_{t_0}^{t}dt_2 s(t_1)s(t_2)f\left (\epsilon_{{\bf k}+e\int_{t_2}^{t_1}dt'{\bf A}(t')/[(t_1-t_2)\hbar c]}-\mu\right )\nonumber\\
&\times&
\exp\left [ \frac{i}{\hbar}\int_{t_2}^{t_1}dt'\left ( \hbar\omega+\mu-\epsilon_{{\bf k}-e\int_{t_2}^{t_1}d\bar t[{\bf A}(t')-{\bf A}(\bar t)]/[(t_1-t_2)\hbar c]}\right )\right ],
\eea
and one can see that the times get entangled in complicated ways that a simple factorization to show it is nonnegative looks to be impossible to carry out. This issue comes from the fact that the average vector potential, averaged over the relative time interval, is subtracted from the vector potential shift in the exponent, and the integral that gives rise to the average value is difficult to deal with. But we can examine more closely some simpler cases to see if we can make progress, or at least understand the complications more clearly.

So, let us look at a constant DC pump, given by $A(t)=-{\bf E}t\theta(t)$, and examine probe functions that are peaked for large positive times. In this case, we can replace $A(t)$ by $-Et$, since its argument is always at large positive times due to the $s(t)$ factors. Then we find
\beq
\tilde P_{\bf k}(t)=\frac{1}{\hbar^2}\int_{t_0}^{t}dt_1 
\int_{t_0}^{t}dt_2 s(t_1)s(t_2)f\left (\epsilon_{{\bf k}-e{\bf E}(t_1+t_2)/(2\hbar c)}-\mu\right )
\exp\left [ \frac{i}{\hbar}\int_{t_2}^{t_1}dt'\left ( \hbar\omega+\mu-\epsilon_{{\bf k}+e{\bf E}[t'-(t_1+t_2)/2]/(\hbar c)}\right )\right ].
\label{eq: p_non}
\eeq
If we work on a $d$-dimensional hypercubic lattice, then $\epsilon_{\bf k}=-2t\sum_{i=1}^d \cos (k_ia)$, so if the field is put in the diagonal direction, and we define $\bar\epsilon_{\bf k}=-2t\sum_{i=1}^d\sin(k_ia)$, then $\epsilon_{{\bf k}+c{\bf E}}=\epsilon_{\bf k}\cos(cE)+\bar\epsilon_{\bf k}\sin(cE)$. The exponential factor in Eq.~(\ref{eq: p_non}) becomes
\begin{equation}
\exp\Bigr [\frac{i}{\hbar}\int_{t_2}^{t_1}dt'\Big\{\hbar\omega+\mu-\epsilon_{\bf k}\cos\left (\frac{eE[t'-(t_1+t_2)/2]}{\hbar c}\right )
-\bar\epsilon_{\bf k}\sin\left (\frac{eE[t'-(t_1+t_2)/2]}{\hbar c}\right )\Big\} \Bigr ].
\end{equation}
\end{widetext}
Performing the integral yields
\begin{equation}
\exp\left [\frac{i}{\hbar}(\hbar\omega+\mu)(t_1-t_2)-\frac{2ic\epsilon_{\bf k}}{eE}\sin\left (\frac{eE(t_1-t_2)}{2\hbar c}\right )\right ].
\end{equation}
In order to prove nonnegativity the same way as we did before, this exponential factor needs to factorize into one function of $t_1$ and one function of $t_2$, with the second function being the complex conjugate of the first function. The term with the $\sin$ in the exponent, does not  appear to factorize this way. In addition, the Fermi-Dirac distribution has an argument that is a complicated combination of $t_1$ and $t_2$ as well. What one can immediately notice is that the term with the Fermi-Dirac distribution is a function of $t_{ave}$  only, while the exponential term is a function of $t_{rel}$ only. The product $s(t_1)s(t_2)$ will generically depend on both average and relative times, but it is an even function with respect to $t_{rel}$. So, if we reorganize the integral into one over the average and relative times, then it has the form of the integrand that depends on average time being nonnegative, while the integrand that depends on relative time is an even nonnegative function in $t_{rel}$ multiplied by the real part of the exponential of $i$ multiplied by an odd function. It is possible that a generalization of Bochner's theorem from spectral analysis~\cite{bochner} would show that such an object is nonnegative for every $t_{ave}$ which would then prove nonnegativity, but it is not obvious to us how this would work. In numerical calculations, we have always found that the gauge-invariant tr-ARPES signal is nonnegative, which makes us believe a proof should be possible. The exposition here clearly shows that if this is the case, then the proof is nontrivial.

\section{Subdominant contributions to the measured photocurrent}

As mentioned earlier, the perturbative analysis for $\mathcal{H}_{\rm probe}$ presented in Ref.~\onlinecite{our-prl} kept and analyzed only the {\em dominant} contribution to the detected photocurrent to second order in $\mathcal{H}_{\rm probe}$, and not the {\em entire} contribution. Indeed, there are two other contributions to the photocurrent $\langle \mathbf{J}_d \rangle(t)$ in Eq.~(\ref{J_d-avg-detail}) that are formally of order $(\mathcal{H}_{\rm probe})^2$ which are   given by
\bea
&-& \frac{1}{(2\hbar)^2} \int_{t_0} ^{t} dt_2 \int_{t_0} ^{t} dt_1 \Big [\langle U(-\infty, t) \mathbf{J}_d  U(t,t_2) \mathcal{H}_{\rm probe}(t_2)  \nonumber\\
&~&\quad\times \; U(t_2,t_1) \mathcal{H}_{\rm probe}(t_1) U(t_1, -\infty) \rangle_{\mathcal{H}} \nonumber\\ &+& \langle U(-\infty, t_2)\mathcal{H}_{\rm probe}(t_2) U(t_2,t_1) \nonumber\\
&~&\quad\times \;    \mathcal{H}_{\rm probe}(t_1)  U(t_1, t)\mathbf{J}_d U(t, -\infty) \rangle_{\mathcal{H}}\Big ]
\label{J_d-avg-additional}
\eea
Correspondingly, there are two additional terms in the expression for $P_{\bf k}(t)$, and the complete expression is given by the following, with all $t$-dependent operators in the Heisenberg picture with respect to $\mathcal{H}_{pump}(t)$:
\begin{widetext}
\bea
P_{\bf k}(t) \equiv  \frac{1}{\hbar^2} \sum_{\nu_1,\nu_1^\prime, \mathbf{k}_{\parallel 1}}  \sum_{\nu_2,\nu_2^\prime, \mathbf{k}_{\parallel 2}} \int_{t_0}^{t} dt_2 \int_{t_0}^{t} dt_1 s(t_2)s(t_1) e^{i\omega_{\mathbf{q}}(t_2-t_1)} M^*_{\mathbf{q}}(\nu_2,\nu_2^\prime; \mathbf{k}_{\parallel 2}; t_2) \, M_{\mathbf{q}}(\nu_1,\nu_1^\prime; \mathbf{k}_{\parallel 1}; t_1) \; \nonumber \\ \times \;\Big [ \langle c^{\dagger}_{\nu_2 \mathbf{k}_{\parallel 2}}(t_2) c_{\nu_2^\prime \mathbf{k}_{\parallel 2}+\mathbf{q}_\parallel}(t_2)
   c^{\dagger}_{\nu^\prime \mathbf{k}_\parallel^\prime}(t) c_{\nu \mathbf{k}_\parallel}(t)  c^{\dagger}_{\nu_1^\prime \mathbf{k}_{\parallel 1}+\mathbf{q}_\parallel}(t_1) c_{\nu_1 \mathbf{k}_{\parallel 1}}(t_1) \rangle_{\mathcal{H}} \; \nonumber\\
    -\frac{1}{2} \langle c^{\dagger}_{\nu^\prime \mathbf{k}_\parallel^\prime}(t) c_{\nu \mathbf{k}_\parallel}(t) c^{\dagger}_{\nu_2 \mathbf{k}_{\parallel 2}}(t_2) c_{\nu_2^\prime \mathbf{k}_{\parallel 2}+\mathbf{q}_\parallel}(t_2)
     c^{\dagger}_{\nu_1^\prime \mathbf{k}_{\parallel 1}+\mathbf{q}_\parallel}(t_1) c_{\nu_1 \mathbf{k}_{\parallel 1}}(t_1) \rangle_{\mathcal{H}} \; \nonumber \\
   -\frac{1}{2} \langle c^{\dagger}_{\nu_2 \mathbf{k}_{\parallel 2}}(t_2) c_{\nu_2^\prime \mathbf{k}_{\parallel 2}+\mathbf{q}_\parallel}(t_2)
     c^{\dagger}_{\nu_1^\prime \mathbf{k}_{\parallel 1}+\mathbf{q}_\parallel}(t_1) c_{\nu_1 \mathbf{k}_{\parallel 1}}(t_1) c^{\dagger}_{\nu^\prime \mathbf{k}_\parallel^\prime}(t) c_{\nu \mathbf{k}_\parallel}(t) \rangle_{\mathcal{H}} \Big  ].
\label{central-result-3}
\eea
\end{widetext}
The first term is what we had earlier, and the other two terms come from the two terms in Eq.~(\ref{J_d-avg-additional}) above. The creation and annihilation operators for the detected (photo)electrons carry the time label $t$ in all of the above terms. Hence it is clear that the physical processes corresponding to the extra terms require the detected electron to come right out of the {\em (time evolved)  initial state, before the photon is absorbed}, which can happen only when that state has an electron excited to a high enough band  (TRL) state that it will come out of the sample. When the pumped system is describable as thermalized with an effective electron temperature $T_e$, the Boltzmann  probability for this is proportional to $\exp[-\epsilon_p / (k_B T_e)]$ where $ \epsilon_p = \epsilon_{\nu^\prime \mathbf{k}^\prime_{\parallel}} - \mu$ is the (excitation) energy of the detected electron measured from the chemical potential (essentially the Fermi level) of the system. This is clearly small as long as $\epsilon_p$ is much larger than $(k_B T_e)$ which is typically the case, and can happen even if the kinetic energy of the detected electrons is not very large, {\it e.~g.} if $\epsilon_p$ is only slightly larger than the work-function, but the latter is much larger than  $(k_B T_e)$. Even when the pumped system is not in a thermal distribution, electrons can only be excited to such higher bands either via a tunneling process, involving a Landau-Zener-like transition which depends on the speed at which the gaps in the band structure are traversed (as they are driven by the pump field) compared to the sizes of the gaps, or via multiphoton absorption processes  requiring multiple dipole transitions; in both cases one expects the population that is excited and the contributions from the neglected terms, to be small.

Within the approximations we have made in this analysis, such as those discussed following Eq.~(\ref{fac-eqn}), which should be quite accurate for the high-energy electrons, and assuming that the pumped system can be approximated as being in quasiequilibrium at an effective electronic temperature $T_e$, one can explicitly evaluate the additional contributions, and verify that they are indeed small. The above Boltzmann factor manifests itself in this case via Fermi-Dirac distribution functions of $\epsilon_p$. There is also a second (Wick) contraction of the average in Eq.~(4) of Ref.~\onlinecite{our-prl} [equivalently, Eq.~(\ref{central-result-2}) above] which was also not considered previously, as it involves Fermi-Dirac functions of $\epsilon_p$, and can similarly be neglected in the context of normal pump-probe photoemission experiments. But needless to say, there might be special experimental circumstances where the extra terms, while small, are measurable, and need to be taken into account, especially when the pump becomes strong and can excite the band electrons higher than expected just from energy conservation stemming from the pump's frequency distribution. Hence we have presented them in detail here.

\section{Conclusions}

In this work, we have discussed a number of issues related to details in the theory of time-resolved and angle-resolved pump/probe photoemission spectroscopy. In particular, we have discussed how one must change the formal results when the pump pulse is present during the same time that the probe pulse is being applied. In this case, one must convert the results for momentum-dependent quantities in a gauge into gauge-invariant quantities, which are the physically measurable results.  Such an approach has already been taken into account in recent work on tr-ARPES in electron-phonon coupled systems~\cite{lex1,prx} and in transient-induced topology changes in graphene~\cite{graphene}. The solution is to replace the momentum-dependent lesser Green's function in the presence of the pump pulse by the so-called gauge-invariant one, which is an {\it ad hoc}  procedure, that is, nevertheless widely used. We discussed the issues behind formulating a fully gauge-invariant theory from the start, but that analysis requires some significant formal development to complete, which is beyond the scope of this work.

Next, we focused on the issue of whether the tr-ARPES signal was nonnegative, which it is required to be since it is interpreted as a probability. The tr-ARPES signal in the vector-potential-only gauge can be easily shown to be nonnegative since it arises directly from the square of a matrix element. Making the transformation to the gauge-invariant Green's function, complicates the analysis significantly because the integrals over time get entangled together, and one cannot see the manifestly nonnegative character of the response.  We investigated this issue more thoroughly by examining the results for a noninteracting single-band model, where one can get an analytic formula for the nonequilibrium Green's function. Even in that case, when one picks a simple constant DC field for the pump, it does not appear obvious at all how to verify the nonegativity. It is likely that the nonnegativity is related to a generalized form of Bochner's theorem from spectral analysis which deals with positive-definite Fourier transforms. 

Finally, we discussed a set of terms that are second-order in the probe Hamiltonian, but were neglected in the previous analysis of tr-ARPES due to their being generically smaller than the terms that we did include.  Those extra terms essentially correspond to the situation where the pump field is responsible for the photoemission, which can occur when it is a large enough amplitude field and has been applied for a long enough time to drive the electrons far from equilibrium. But, we expect that even in those cases, the signal will be dominated by the term that we did keep, and these extra terms will provide only a small correction. It would be interesting to find experimental circumstances where those types of terms can dominate the response.

\acknowledgments
We thank Andrij Shvaika and Oleg Matveev for drawing our attention to the fact that we had dropped the extra terms of order $(\mathcal{H}_{\rm probe})^2$ without discussion. 
This work was supported by the Department of Energy,
Office of Basic Energy Sciences, Division of Materials
Sciences and Engineering (DMSE) under Contracts
No. DE-AC02-76SF00515 (Stanford/SIMES), No. DE-FG02-08ER46542 (Georgetown), and No. DE-SC0007091
(for the collaboration). 
We also acknowledge support of the Indo-US Science and Technology Forum under a center grant numbered JC-18-2009 for supporting the Indo-US collaboration. HRK acknowledges support from the DST, India, and J.K.F. also acknowledges support by the McDevitt
bequest at Georgetown.

%\end{itemize}

\end{document}